# Reduction of prolonged excessive pressure in seated persons with paraplegia using wireless lingual tactile feedback: a randomized controlled trial

A. Moreau-Gaudry, O. Chenu, M. V. Dang, J-L. Bosson, M. Hommel, J. Demongeot, F. Cannard, B. Diot, A. Prince, C. Hughes, N. Vuillerme, Y. Payan

**Abstract:** Pressure ulcers (PU) are known to be a high-cost disease with a risk of severe morbidity. This work evaluates a new clinical strategy based on an innovative medical device (Tongue Display Unit - TDU) that implements perceptive supplementation in order to reduce prolonged excessive pressure, recognized as one of the main causes of PU. A randomized, controlled, parallel-group trial was carried out with 12 subjects with spinal cord injuries (SCI). Subjects were assigned to the control (without TDU, n=6) or intervention (with TDU, n=5) group. Each subject took part in two sessions, during which the subject, seated on a pressure map sensor, watched a movie for one hour. The TDU was activated during the second session of the intervention group. Intention-to-treat analysis showed that the improvement in adequate weight shifting between the two sessions was higher in the intervention group (0.84 [0.24; 0.89]) than in the control group (0.01 [-0.01; 0.09]; p=0.004) and that the ratio of prolonged excessive pressure between the two sessions was lower in the intervention group (0.74 [0.37; 1.92]) than in the control group (1.72 [1.32; 2.56]; p=0.06). The pressure map sensor was evaluated as being convenient for use in daily life, however this was not the case for the TDU. This work shows that persons with SCI could benefit from a system based on perceptive supplementation that alerts and guides the user on how to adapt their posture in order to reduce prolonged excessive pressure, one of the main causes of PU.

## I. INTRODUCTION

PRESSURE ulcers (PU) are one of the most common long-term secondary complications in persons with spinal cord injury (SCI). Their estimated prevalence varies between 17 % and 33 % of community resident persons with SCI [1]. PU require burdensome care and alter quality of life through pain, need of assistance, wound infections and long hospitalizations. Treating one pressure ulcer costs between 2,000 and 20,000 US$ depending on severity, and the added overall cost of PU ranges between 2.1 and 3.8 billion US$ for the United Kingdom in the year 2000, i.e. 4 % of the National Health System budget [2]. PU are thus considered to be a major public health issue, and their prevention constitutes an important research area.

Factors influencing the occurrence of PU are multiple, including unrelieved pressure, shear, friction, moisture, poor nutrition, immobility, psychological distress and social difficulties. Prolonged excessive pressure near bony prominences is considered in particular to be a key element for

Manuscript submitted December 2017.

This research was supported by a grant from the Délégation Régionale à la Recherche Clinique (DRRC) of Grenoble Alpes University Hospital (reference PL/FM/BM DRRC 2006).

A. Moreau-Gaudry is with the INSERM CIC 1406, Grenoble, France, the Grenoble Alpes University Hospital, Grenoble, France, the Université Grenoble Alpes, TIMC-IMAG, Grenoble, France and the CNRS, TIMC-IMAG, Grenoble, France. O. Chenu was with the Université Grenoble Alpes, TIMC-IMAG, Grenoble, France and is now with the TexiSense S.A. company, Montceau-les-mines, France (correspondence e-mail: olivier.chenu@texisense.com). M.V. Dang was with the INSERM CIC 1406, Grenoble, France, the Grenoble Alpes University Hospital, Grenoble, France, the Université Grenoble Alpes, TIMC-IMAG, Grenoble, France and the CNRS, TIMC-IMAG, Grenoble, France. J-L. Bosson is with the INSERM CIC 1406, Grenoble, France, the Grenoble Alpes University Hospital, Grenoble, France, the Université Grenoble Alpes, TIMC-IMAG, Grenoble, France and the CNRS, TIMC-IMAG, Grenoble, France. M. Hommel was with the INSERM CIC 1406, Grenoble, France and the Grenoble Alpes University Hospital, Grenoble, France. J. Demongeot was with the Université Grenoble Alpes, TIMC-IMAG, Grenoble, France and the Institut Universitaire de France, Paris, France. F. Cannard is with the TexiSense S.A. company, Montceau-les-mines, France. B. Diot is with the IDS S.A. company, Montceau-les-mines, France and the Université Grenoble Alpes, AGEIS, Grenoble, France. A. Prince is with Clinique du Grésivaudan, La Tronche, France. C. Hughes is with the INSERM CIC 1406, Grenoble, France, the Université Grenoble Alpes, CIC 1406, Grenoble, France and the Grenoble Alpes University Hospital, CIC 1406, Grenoble, France. N. Vuillerme is with the Université Grenoble Alpes, AGEIS, Grenoble, France and the Institut Universitaire de France, Paris, France. Y. Payan is with the Université Grenoble Alpes, TIMC-IMAG, Grenoble, France and the CNRS, TIMC-IMAG, Grenoble, France.

the onset of muscular tissue necrosis leading to deep PU [3]. During long sitting periods, the two main factors that increase the risk of deep PU for persons with SCI are impaired mobility and a sensitivity deficit in the lower limbs that alters their capacity to feel excessive pressure in soft tissues under bony prominences. As a consequence, their ability to shift posture according to an alert signal perceived consciously or unconsciously, is reduced in comparison to healthy persons. Preventive strategies thus recommend that persons with SCI be trained to shift their weight every 15 to 30 minutes when seated [4]. However, this trained behavior appears to fade away in most persons once they return home after a rehabilitation program. In order to palliate the deficient weight shifting behavior, it has been proposed to create pressure changes using innovative medical devices, such as alternating pressure cushions [5], off-loading [6] or dynamic sitting wheelchairs [7].

In this work, we evaluate the principle of perceptive supplementation in the target population of persons with SCI. This principle is based on the supplementation of the sensitivity deficit, relative to the information concerning the prolonged pressure applied on the buttock area, by transmission through another functional sensory channel. The validity of the principle has been proven for blind persons in the approach designed by P. Bach-y-Rita [8]: the deficient vision of blind persons was supplemented by electrodes placed on their backs that produced tactile stimuli according to the visual scene in front of them. After an adaptation period, the blind person forgot the tactile stimuli and perceived objects to be at their actual place. Our hypothesis is that persons with SCI could also use a supplementary sensory input in order to augment the deficient sensitivity of their buttocks, and could thus adapt their weight shifting to the actual risk in their gluteal soft tissues. In this way, the person with SCI would maintain control of their dynamic sitting behavior rather than having a weight shifting pattern imposed upon them.

Although several modalities are possible for sensory input, we chose to use the tongue surface for this work. This choice was motivated by several reasons: firstly, Bach-y-Rita, a reference in the field of perceptive supplementation, obtained better perception results using electrical stimulation of the tongue than tactile systems [9, 10]. That the system is now commercialized as an FDA approved vision aid (BrainPort, Wicab, Inc.) and our own previous experience [11-13] demonstrate the feasibility of using the tongue for sensory input. Secondly, the tongue is the organ that possesses the greatest density of tactile receptors and its discriminative tactile capacity exceeds even that of the fingertips. Furthermore, the tongues moisture ensures good electrical contact and because of the protected environment of the mouth, the hydration and hence electrical properties of the tongue are more consistent than those of the skin, meaning that simpler voltage-control circuitry can be used [14]. Moreover, in persons with SCI, the tongue is one of the few organs to remain entirely free of deficiencies. A final important aspect concerns the stigmatization associated with the use of visible assistive devices. Sensory input via the tongue can be made such that it is not externally visible, therefore making the device more cosmetically acceptable.

To activate the lingual sensory channel, a person with SCI places a wireless lingual device (the Tongue Display Unit - TDU) in their mouth; the device delivers an electrostimulation to the tongue according to input from a pressure map sensor installed on the seat of the person's wheelchair. The underlying idea is to transfer via the tongue the "pain" that the person does not adequately perceive from his/her buttocks because of the SCI. Such a TDU device has already been described in previous studies performed on healthy subjects; these studies have shown that healthy subjects are able to successfully use the device in order to adequately shift their pressure load [15]. Another study has shown that a person with paraplegia could benefit from the TDU as a supplementary sensory input [16].

Many studies have evaluated weight shifting behavior using subjects with no sensory deficiency or mobility limitation. We have chosen to evaluate the clinical relevance of our proposed approach using persons with SCI. To the best of our knowledge, this study is the first time that such an evaluation has been carried out.

Three elements were assessed by this clinical evaluation: (i) the effect of the TDU guided stimulations on the postural behavior of the subjects, (ii) the efficiency of the proposed strategy in terms of the reduction of prolonged excessive pressure at the seat/skin interface and (iii) the usability of the developed medical device. To evaluate these three complementary elements, a prospective, randomized, controlled clinical trial was designed in order to obtain a high level of clinical evidence in the target population (ClinicalTrials.gov identifier: NCT00429013 - Pressure Ulcer Formation Prevention in Paraplegics Using Computer and Sensory Substitution Via the Tongue).

## II. MATERIALS AND METHODS

### A. Study design

This was an open, monocentric, randomized, parallel-group study, conducted at the Clinical Investigation Centre – Innovative Technology (CIC-IT) of the Grenoble Alpes University Hospital.

### B. Study subjects

Eligible subjects were persons aged 18 years or over with SCI. The exclusion criteria were as follows: pregnancy, lactation, known cognitive pathologies, persons under legal protection according to French regulation related to biomedical research, persons with pressure ulcers, persons with an intolerance to wearing a palate prosthesis, persons with buccal pathologies and persons unable to autonomously manipulate the palate prosthesis.

### C. Materials

The medical device consisted of 3 parts (illustrated in Fig. 1): (i) the pressure map sensor is composed of 32x32 sensors, each of approximate dimension 1 cm², that can measure pressure from 0 to 200 mmHg, with a precision less than 1 mmHg. The data acquisition frequency was approximately 10 Hz. (ii) The

TDU is a wireless electrotactile device that consists of an orthodontic palate including a battery, a radio chip for communication and a 6x6 electrotactile matrix. The electric signal conducted by the electrodes is similar to the one designed by Tyler et al [9], except that the maximum current is 5V. The current is limited in order to improve safety and to simplify the design of such a small device. It was previously verified that 5V was sufficient for users to perceive the electrostimulation of the tongue. (iii) A laptop containing in-house written software that connects the pressure map sensor and the TDU.

*D. Methods*

Eligibility assessment, information delivery before the subject gave written informed consent and the inclusion of eligible subjects was performed by the principal investigator of the study. Subjects were randomly assigned to the intervention group (with the TDU) or the control group (without the TDU). The random assignment was carried out using a computer generated list of random numbers.

Each subject took part in two sessions, S1 and S2. S2 took place one week after S1. During each session, the subject was seated in their own wheelchair on a flexible pressure map sensor that recorded the pressure applied at their buttock area. Each session lasted for one hour during which the subjects watched a movie. Before S1, the subject chose a movie to watch from a selection. Before S2, the subject chose to either continue watching the same movie or chose a different movie to watch from the selection.

Before each session, a calibration was performed. For the calibration, the subject was asked to position himself in nine different postures (front-left, front, front-right, left, center, right, rear-left, rear and rear-right). For each posture, the system recorded the corresponding pressure map (referred to as the reference maps). During the following session, the current posture of the subject at any given time was determined by comparing the current pressure map with the nine reference maps. The most similar reference map was considered to be the current posture. The nine postures were therefore defined individually for each patient, depending on the ability of the subject to adopt each of the nine postures.

The sessions were carried out as follows: the session was started (t=0) and the pressure data was then read in real time. At t=t1 (60 +/- 5 s), the current posture of the subject was estimated and the "cumulative excessive pressure map" (CEPM) from t=0 to t=t1 was calculated. The CEPM enables the area that has been the most exposed to excessive pressure during the previous 60 s to be identified. Considering the estimated current posture and the calculated CEPM, a target posture was determined that most decreases the risk arising from excessive pressure. Then, from t=t1 to t=t1+10 s, the algorithm evaluated if the target posture had been achieved: if so the weight shift was considered to be adequate. Fig. 2 illustrates the described process of adequate weight shifting. The described process was repeated for the duration of the session.

For S1 and S2 of the control group and for S1 of the intervention group, no directional instruction or any feedback of any kind was sent to the subject. For subjects assigned to the intervention group, the TDU was activated during the second session (S2). The previously described process was modified as follows: at t=60 s, a directional instruction on how to change the current posture to the target posture was sent to the subject via the TDU. The TDU signal was stopped as soon as the target posture was achieved or after 10 s. In the latter case (target posture not achieved after 10 s), the signal was stopped after 10 s as it was considered that either the signal was not perceived or the target posture was too difficult to achieve. The described process was repeated for the duration of the second session. The alert frequency was defined as 60 s as it has been estimated that able bodied office workers change seated position approximately once per minute [17].

The CEPM was computed as follows: the mean of the pressure maps acquired since t=0 was calculated; from this mean pressure map, all values below an "unsafe" pressure threshold – Prisk – were disregarded and set to zero. The threshold, Prisk, was defined as 100 mmHg, which is in the range of physiological arterial blood pressure.

The target posture was calculated as follows: from the estimated current posture at t=t1, four easily achievable target postures were calculated (by a left, right, front or rear movement of the trunk). For each of these four achievable postures, the pressure decrease of the CEPM if the subject adopted the posture was computed. The posture that most decreased the pressure in the exposed area was considered the target posture.

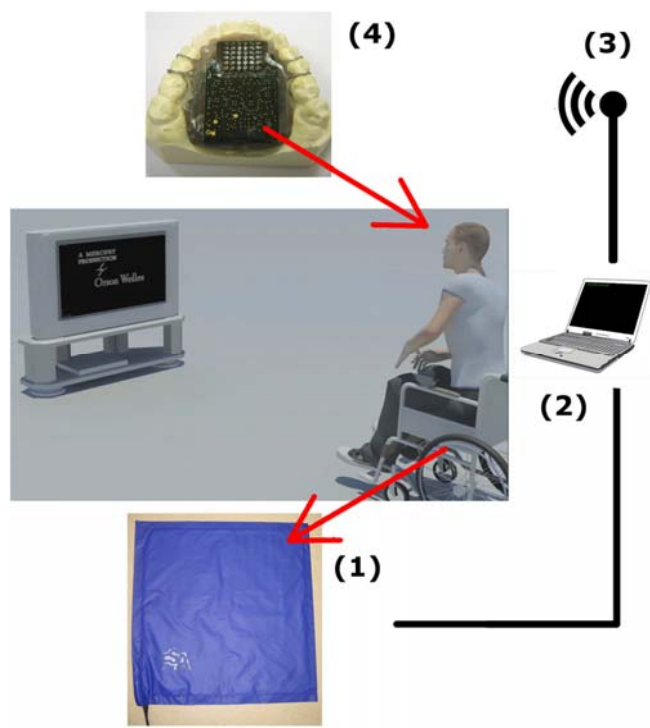

Fig. 1. Materials used in the study: while the subject is watching a movie, the pressure map sensor measures the pressure data (1) and sends the data to the laptop (2), which analyses the pressure and - during the second session of the intervention group – initiates every 60 +/- 5 s a wireless communication (3) with the TDU (4) which is placed inside the subject's mouth, and which indicates to the subject to move in a given direction by means of an electrostimulation of the tongue.

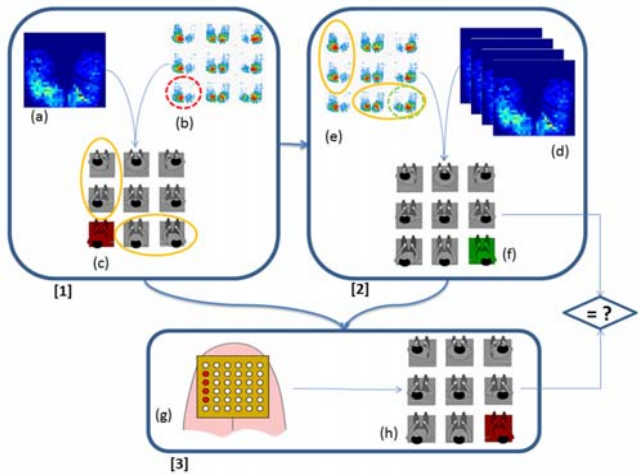

Fig. 2. This figure illustrates the process of adequate weight shifting, which consists of the following three main steps: [1] determination of the current posture of the subject and the four easily achievable target postures: the current pressure map of the subject (a) is compared to the reference maps (b). The most similar reference map, here the posterior-left (red dashed line circle), is considered to be the current posture (c, red). From this posture, the four postures achievable in a single movement (in antero-posterior or medio-lateral direction) are identified (orange ellipses). [2] determination of the target posture: the CEPM enables the most exposed area during the last 60 +/- 5 s to be identified (d); the most exposed area is compared to the four achievable postures maps (e, orange ellipses) identified in step [1]; the map that most decreases the pressure in the exposed area, here the posterior-right (green dashed line circle), is considered the target posture (f, green). [3] electrostimulation and evaluation of the final posture: knowing the current and the target postures, a directional signal is sent by the TDU to the tongue (g) according to the principle of avoidance. In this example, the subject has to move from posterior-left to posterior right; as he must move to the right, the left part of the TDU is activated. The final posture of the subject is evaluated (h) in the same way as in step [1]. In this example, the final posture is posterior-right (h, red), this corresponds to the target posture (f, green), we therefore consider the movement to have been correctly executed and the weight shift is considered adequate.

The directional instruction, sent by the TDU to the tongue, was indicated according to the well-known principle of "avoidance", that is, by activating the electrodes on the opposite side of the target posture [18]. The signal is an alert signal that enables the subject to react to reduce the "pain" and thus avoid a danger (i.e. the excessive pressure). For example, activating the left side of the TDU means "avoid this and go towards the right side". This concept is illustrated in Fig. 2.

*E. Definition of R - the estimation of the risk of developing a PU*

In order to quantify the estimation of the risk of developing a PU, a risk index, R, was defined. Though many factors have been identified that contribute to the risk of developing a PU when a pressure is applied on soft tissue, the two main factors are the level of applied pressure and the duration of the applied pressure. An "unsafe" level of pressure may lead to decreased tissue perfusion. If such a decreased tissue perfusion persists for an extended amount of time, it may create an ischemia that can lead to tissue necrosis if left untreated. The risk index R is therefore defined by considering the "interaction" of these two factors: the higher the "unsafe" pressure level is, the worse the tissue perfusion may be. Furthermore, the longer the "unsafe" pressure is continuously applied, the greater the risk of being in conditions that may promote the development of a PU.

The risk index R is therefore defined as the amount of time during which an inadequate tissue perfusion may be encountered, with the amount of time weighted by the level of inadequate perfusion. Inadequate perfusion is correlated with excessive pressure, i.e. the pressure difference between the current level of applied pressure and an "unsafe" pressure threshold (Prisk, previously defined at 100 mmHg).

The calculation of R for one sensor during a session is as follows (illustrated in Fig. 3): (A) the pressure $P(t)$ applied to the sensor is measured with respect to time; (B) the indicator function $1[P(t) \geq Prisk]$ is determined; (C) the instantaneous excessive pressure $IEP(t)$, defined as $(P(t)-Prisk)1[P(t) \geq Prisk]$, is then calculated; (D) the instantaneous time-cumulated excessive pressure $IC(t)$, defined as $IEP(t)$ cumulated over each period of time for which $P(t) \geq Prisk$, is determined; (E) the instantaneous risk $IR(t)$, defined as the instantaneous time-cumulated excessive pressure normalized by the mean excessive pressure, is then evaluated. The normalization is applied in order to take into account the variability of weight between subjects. Finally, the risk index R associated with the sensor, defined as the mean of IR(t) over the session, is calculated. The risk index of the entire pressure map sensor is defined as the sum of the risk indices calculated for each sensor.

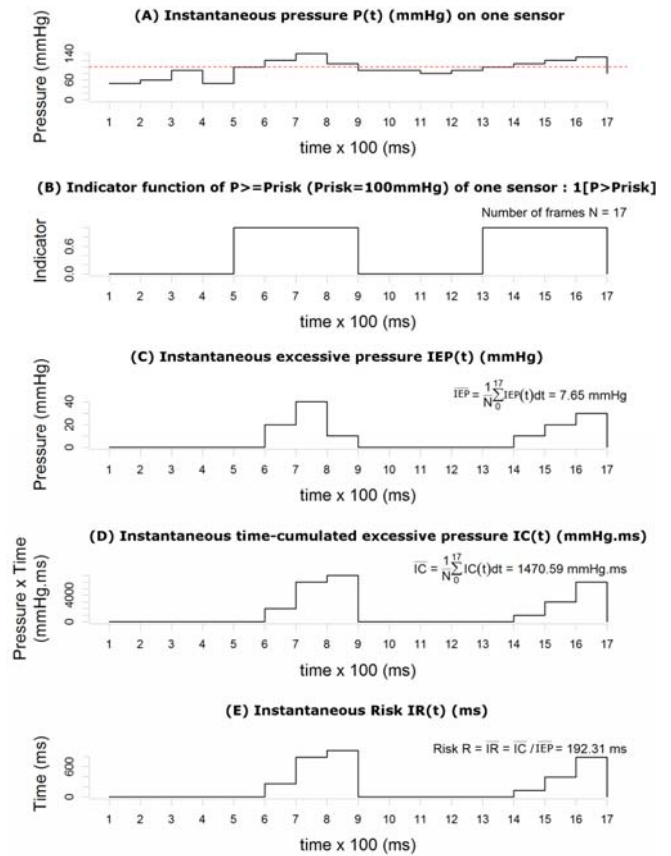

Fig. 3. The calculation of the risk index R, for one sensor during a session, is as follows, using a 1.7 s example for illustration: (A) the instantaneous pressure $P(t)$ measured by one sensor of the pressure map sensor; the red dashed line indicates the defined "unsafe" threshold, Prisk = 100 mmHg, above which pressures are considered to be potentially dangerous; (B) the indicator function $1[P \geq Prisk]$ is computed that indicates the temporal ranges of potentially dangerous pressures; (C) the instantaneous excessive pressure, $IEP(t) = 1[P \geq Prisk]*(P(t)-Prisk)$, is then determined; (D) the instantaneous time-cumulated excessive pressure, $IC(t)$, is calculated, which represents the accumulation of dangerous instantaneous excessive pressures; (E) the instantaneous risk, $IR(t)$, is then calculated by dividing $IC(t)$ by the mean excessive pressure. The risk index R of the sensor is defined as the mean of $IR(t)$ over the session. The risk index of the entire pressure map sensor is defined as the sum of the risk indices calculated for each sensor.

*F. Evaluation of the usability of the medical device*

The acceptability of the medical device for use in daily life was evaluated by the subjects after the second session by means of a questionnaire. The pressure map sensor and TDU were evaluated separately, according to the facility of set-up, removal and usage and the comfort and psychological impact of usage. Each aspect was evaluated using a scale that varied from 0 (acceptable) to 10 (unacceptable) for use in daily life.

*G. Outcomes*

The main outcome was to evaluate the evolution in the proportion of adequate weight shifts between S1 and S2 and to compare the evolution between the intervention group and the control group.

Two secondary outcomes were evaluated: the first outcome was to evaluate the evolution of the estimation of the risk of developing a PU between S1 and S2 and to compare between the intervention group and the control group. The second outcome was to evaluate the acceptability of the medical device for use in daily life.

*H. Statistical analysis*

The statistical analysis was performed at the CIC-IT of the Grenoble-Alpes University Hospital. Continuous variables were expressed using medians and interquartile ranges [Q1; Q3] and compared using Wilcoxon tests (W). Data were analyzed according to both the worst-case scenario intention-to-treat (ITT) and per-protocol (PP) principles. The global significance threshold was set at 0.05 for the main and the secondary outcomes.

An interim analysis of the main outcome was planned in order to prematurely stop the study if possible to avoid unnecessarily recruiting subjects with SCI. To prevent the risk inflation of the alpha level, the significance threshold for this interim analysis was set at 0.01 (p<0.01).

The sample size, taking into account the interim analysis, was calculated from modelling the expected results and estimated to be 12 subjects per group. The interim analysis was therefore planned after five subjects per group had been recruited.

The statistical analysis was performed using GNU-R software, version 3.0.3 "Warm puppy" [19].

### III. RESULTS

*A. Study subjects*

Twelve subjects were recruited from September 2006 to October 2008. However, although subject 10 was initially included, she was subsequently excluded due to an unexpected pregnancy. This subject did not take part in any session. Thus, after randomization, six subjects were assigned to the control group (without the TDU) and five subjects were assigned to the intervention group (with the TDU). Concerning the control group, the data of one subject (the first subject included in the study) was not analyzed because of a software failure: the nine calibration reference maps necessary to determine the posture of the subject during the subsequent session were not recorded. Concerning the intervention group, the data of one subject (the last subject included in the study) was not analyzed because of an irreversible material failure in the TDU (non-functioning wireless module, identified at the beginning of the first session). A flow chart summarizing the subjects included in the study is presented in Fig. 4.

An interim analysis had been planned after the inclusion of ten effectively analyzable subjects, in order to stop the study early if the main outcome had already been proved. However, because of the irreversible failure of the TDU and the excessive cost necessary to repair it, it was decided to conduct the statistical analysis on the nine PP subjects and the eleven ITT subjects that had been included. As the main outcome was proved (p<0.01), the steering committee decided to stop the study.

Table I resumes the clinical characteristics of the subjects on which the PP analysis was performed: age, gender, weight, height, body mass index (BMI), ASIA (American Spinal Injury Association) score used to characterize the level of neurological deficiency for persons with SCI [20, 21] and the number of subjects with post-traumatic PU.

*B. Main outcome: adequate weight shifting*

The main outcome was to evaluate the evolution in the proportion of adequate weight shifts (a shift of the body in order to relieve locally prolonged high pressure) and to compare between the intervention group and the control group. The results are presented in Fig. 5 and Table II. For S1 and S2 of the control group and S1 of the intervention group, no directional instruction or any feedback of any kind was sent to the subject. It was therefore expected that the proportion of adequate weight shifts would be low (control group: median p1 = 0.09, median p2 = 0.09; intervention group: median p1 = 0.04). These measurements constitute a reference value to which the proportion of adequate weight shifts in the intervention group using the activated TDU could be compared (intervention group: median p2 = 0.89). Between the first and second session, the proportion of adequate shifts was increased more by every subject in the intervention group than the subjects in the control group. The one-sided Wilcoxon test concludes that the intervention group had significantly greater values than the control group (ITT analysis: p=0.004, W=30; PP analysis: p=0.008, W=20). The magnitude of increase was greater than + 80 % for three of the four subjects in the intervention group, while in the control group it was less than + 11 %. Thus, once equipped with an activated TDU, a person with SCI shifted his or her weight in an adequate way more than 8 times out of 10.

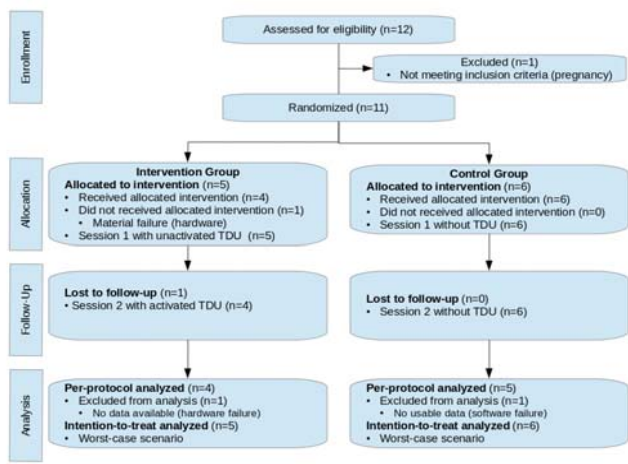

Fig. 4. Flow chart that details the twelve subjects recruited for the study. After inclusion, subject 10 was excluded because of an unexpected pregnancy (exclusion criteria). Six subjects were assigned to the control group and took part in two identical sessions separated by one week. During each session, the pressure applied at the buttock area of the subject was recorded during one hour. Five subjects were assigned to the intervention group and followed the same experimental protocol, but were equipped with a TDU that was activated during the second session.

*C. Secondary outcome: estimation of the risk of developing a PU*

The secondary outcome was to evaluate the evolution of an estimation of the risk of developing a PU and to compare between the intervention group and the control group. The originally defined significance threshold for the secondary outcome (p<0.05) was not adjusted for multiple analyses as only one analysis of the data was carried out. The secondary outcome results are presented in Fig. 6 and Table III. In Fig. 6, the initial distribution of risk index R1 (risk index R during

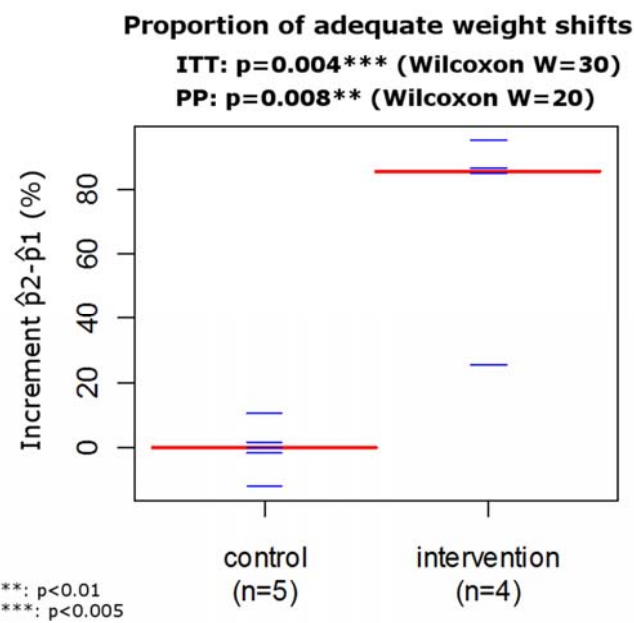

Fig. 5. Evolution in the proportion of adequate weight shifts. Short lines indicate individual results. Long lines indicate the median within each group. The p-value is for a one sided Wilcoxon test between the intervention and control group. ** signifies a p-value < 0.01 and *** signifies a p-value < 0.005.

TABLE I
CLINICAL CHARACTERISTICS OF THE SUBJECTS OF EACH GROUP

|  | Control group (n=5) | Intervention group (n=4) |
|---|---|---|
| Age (years) | 29 [25; 46] | 30 [26; 37] |
| Gender (female) | 1 (20%) | 0 (0%) |
| Weight (kg) | 71.00 [58.50; 83.75] * | 66.00 [61.75; 67.75] |
| Height (cm) | 180.00 [176.00; 183.00] | 175.50 [173.25; 179.00] |
| BMI (kg/m²) | 21.57 [19.02; 24.67] * | 20.40 [19.50; 21.20] |
| ASIA score | 4A (80%); 1B (20%) | 3A (75%); 1D (25%) |
| Nb. subjects with post-traumatic PU | 2 (40%) | 1 (25%) |

Data are expressed as median [Q1; Q3] or numbers (%); * = weight and BMI were calculated for only four subjects because one subject refused to be weighed; ASIA scores range from A (no sensory or motor function preserved in the sacral segments) to E (normal) [20, 21].

TABLE II
EVOLUTION IN THE PROPORTION OF ADEQUATE WEIGHT SHIFTS

|  | Subject | p1 | p2 | Δp | median [Q1; Q3] |
|---|---|---|---|---|---|
| Control group | 1 (ITT) | NA | NA | 0.11 | |
|  | 2 | 0.00 | 0.11 | 0.11 | 0.00 [-0.02; 0.02] |
|  | 6 | 0.11 | 0.13 | 0.02 | |
|  | 8 | 0.09 | 0.07 | -0.02 | *0.01 [-0.01; 0.09] (ITT)* |
|  | 9 | 0.22 | 0.09 | -0.13 | |
|  | 11 | 0.00 | 0.00 | 0.00 | |
| Intervention group | 3 | 0.16 | 0.40 | 0.24 | |
|  | 4 | 0.00 | 0.84 | 0.84 | 0.86 [0.69; 0.90] |
|  | 5 | 0.04 | 0.98 | 0.94 | |
|  | 7 | 0.04 | 0.93 | 0.89 | *0.84 [0.24; 0.89] (ITT)* |
|  | 12 (ITT) | NA | NA | 0.24 | |

p1 = proportion of adequate weight shifts for S1; p2 = proportion of adequate weight shifts for S2; Δp = p2 – p1; the use of italics indicates that missing data has been coded according to the worst-case scenario.

TABLE III
EVOLUTION OF THE ESTIMATION OF THE RISK OF DEVELOPING A PU

|  | Subject | R1 | R2 | R2/R1 | median [Q1; Q3] |
|---|---|---|---|---|---|
| Control group | 1 (ITT) | NA | NA | 1.27 | |
|  | 2 | 2.5 | 3.5 | 1.40 | R1: 2.5 [2.5; 8.2] |
|  | 6 | 8.2 | 16.7 | 2.04 | R2: 6.4 [3.5; 11.7] |
|  | 8 | 9.2 | 11.7 | 1.27 | R2/R1: 2.04 [1.40; 2.08] |
|  | 9 | 1.2 | 2.5 | 2.08 | *R2/R1: 1.72 [1.32; 2.56]* |
|  | 11 | 2.5 | 6.4 | 2.56 | |
| Intervention group | 3 | 1.3 | 2.5 | 1.92 | |
|  | 4 | 3.8 | 2.8 | 0.74 | R1: 2.9 [1.8; 5.2] |
|  | 5 | 9.5 | 1.8 | 0.19 | R2: 2.2 [1.5; 2.6] |
|  | 7 | 1.9 | 0.7 | 0.37 | R2/R1: 0.55 [0.32; 1.03] |
|  | 12 (ITT) | NA | NA | 1.92 | *R2/R1: 0.74 [0.37; 1.92]* |

R1 = estimation of the risk of developing PU during S1; R2 = estimation of the risk of developing PU during S2; R2/R1 = evolution of the estimation of the risk between S1 and S2; the use of italics indicates that missing data has been coded according to the worst-case scenario.

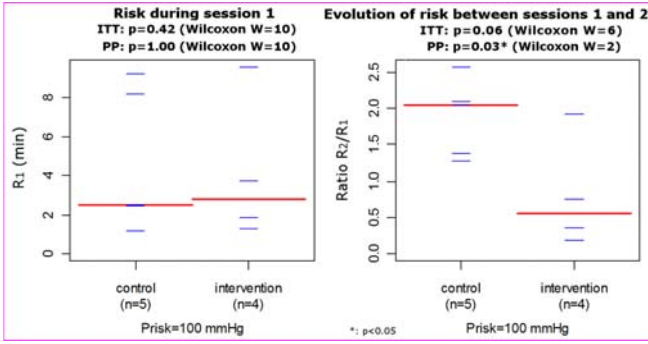

Fig. 6. Evolution of the estimation of the risk of developing a PU between the two sessions. Left panel: distribution of the risk index during session 1 (R1). Right panel: risk index evolution R2/R1 (where R2 is the risk index during session 2). Short lines indicate individual results. Long lines indicate the median within each group. The p-value is for a Wilcoxon test between the intervention and control group (two-sided test on the initial value, one-sided test on the evolution). * signifies a p-value < 0.05.

session 1) in the control and intervention group is shown in the left panel and the risk index evolution R2/R1 (where R2 is the risk index R during session 2) is shown in the right panel. According to the one-sided Wilcoxon test, although no statistical difference is observed between the control and intervention group during the first session (left panel), according to the PP analysis, the intervention group has a statistically significant (p<0.05) lower risk index ratio than the control group (right panel, ITT: p=0.06, W=6; PP: p=0.03, W=2). The absence of statistical significance in the ITT analysis results is considered in the discussion. Considering the PP analysis results, the risk index was decreased more for subjects equipped with lingual feedback than for subjects without lingual feedback. It may be noted that between the first and the second session, the risk index R was on average halved in the intervention group (PP: median [Q1; Q3], R2/R1 =0.55 [0.32; 1.03]), whereas it doubled in the control group (PP: R2/R1 =2.04 [1.4; 2.08]). This result was not due to differing initial risk index values (R1), as there was no significant difference in the R1 values of the two groups (Fig. 6, left panel, Wilcoxon test p=1).

### D. Detailed analysis of individual subjects

It can be seen in the right panel of Fig. 6 and in Table III, that the risk index of subject 3 increased during the second session rather than decreased as was the case for each of the other subjects of the intervention group. For subject 3, the risk index almost doubled: R2/R1= 2.5/1.3 = 1.92. To better understand this "atypical" weight shifting behavior, Fig. 7 shows the timeline of instantaneous risk IR(t), not only for this subject, but also for two other subjects: one is representative of what is observed in the control group (subject 2, top row), and one is representative of what is observed in the intervention group (subject 7, middle row). The pressure map readings for the three subjects during S1 and S2, used to calculate the timelines of instantaneous risk IR(t) shown in Fig. 7, are shown in the video attached to this article.

During the first session, subject 2 from the control group appears to frequently shift weight until t=6 min (zone A: instantaneous risk IR(t) kept near zero). From t=6 min to t=11 min, the subject appears to not move and IR(t) increases linearly during the five minutes (zone B). Then, a partially effective weight shift occurs (point C: IR(t) drops from 6 min to 3 min); and then a completely effective weight shift is observed at t=14 min (point D: IR(t) drops to zero). During this first session, subject 2 appears to produce at least partial weight shifts every five minutes or less. During the second session, the same subject displays two long periods of 12 minutes without any weight shift. The risk index accordingly increases from R1=2.5 min for the first session to R2=3.5 min for the second session. Between the two sessions, a similar risk index increase is observed for every subject of the control group (R2/R1 > 1, cf. Fig. 6).

In contrast, it is apparent from the middle plots of Fig. 7 that subject 7 from the intervention group shifted weight much more often and effectively during the second session than during the first session. In fact, during the second session, this subject performed weight shifting according to the process defined using the activated TDU (that is, every minute and in accordance with the indicated direction) without any evidence of fatigue. For subject 7, the risk index thus decreased from R1=1.9 min to R2=0.7 min between the two sessions.

Subject 3 appeared to shift weight quite frequently during the first session: periods without perceptible weight shifts lasted mostly 1-2 min and were always less than 4 min. During the second session, however, the instantaneous risk IR(t) steadily increases during two long periods: from t=18 min to t=30 min, after which it then drops to zero, and then from t=30 min until the end of the session. When examining the contribution of each of the 32x32 sensors of the pressure map sensor, a single sensor

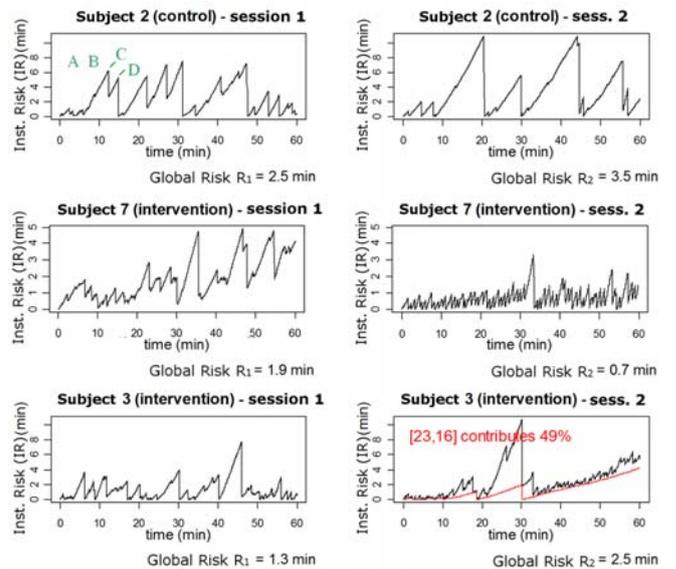

Fig. 7. Timeline of instantaneous risk IR(t) for subjects 2, 7 and 3 during S1 and S2. Subject 2 and 7 are representative of observations in, respectively, the control and intervention groups. Timeline reading for subject 2, first session: frequent and effective shifts (A), no movement (B), partially effective shift (C), completely effective shift (D). Subject 3 differs from the other participants of the intervention group in that the risk index increases from session 1 to session 2 (R1=1.3 min, R2=2.5 min) instead of decreases. For this subject, during S2, one single sensor at u0={23,16} contributed to 49% of the risk index measured over the 32x32 pressure map sensor; the timeline of cumulative excessive pressure R(u0,t) is plotted in red for this sensor.

u0 contributed to 49% of the risk index measured during the second session. The sensor u0 was located at coordinates {23,16} of the pressure map, which corresponded to approximately the center of the left buttock of the subject. The sensor is indicated by a red circle in the attached video. The timeline $IR(u_0,t)$ of this sensor u0 (red line in the bottom right plot of Fig. 7) accounts for the majority of the cumulative excessive pressure measured during the longest steady increase of $IR(t)$. The peaks could be interpreted as weight shifts that relieved the rest of the seat surface, except this spot.

*E. Secondary outcome: usability of the medical device*

The evaluation by the subject of the acceptability of the medical device for use in daily life was an important element to consider. The two parts of the device, the pressure map sensor and the TDU, were evaluated separately and the results are presented in Table IVa and Table IVb respectively.

The pressure map sensor was evaluated according to the facility of set-up (SetUp), the facility of usage (FacilityUse), the comfort of usage (ComfUse) and the psychological impact of usage (PsyImp) of the pressure map sensor in daily life. The TDU was evaluated according to the facility of set-up (SetUp), the facility of removal (Remove), the loss of tongue sensitivity (LossSens), the comfort of usage (ComfUse) and the psychological impact of usage (PsyImp) of the TDU in daily life. Each aspect was evaluated from 0 (acceptable) to 10 (unacceptable) for use in daily life.

The pressure map sensor was evaluated as being very easy to set-up and use, very comfortable to use and psychologically fairly convenient (median inconvenience 3). Conversely, the TDU was evaluated as being very inconvenient (median inconvenience 10), though very easy to set-up and remove (median difficulty 1).

## IV. DISCUSSION

The objective of this study was to evaluate the feasibility and the benefit of using a system based on perceptive supplementation in order to reduce prolonged excessive pressure in the buttock area in persons with SCI. The long-term goal of this study is to help such persons prevent the formation of PU caused, in particular, by impaired mobility and a sensitivity deficit.

The innovative medical device used in this study was developed at TIMC-IMAG laboratory (Grenoble-Alpes University, France). As the device is not CE-marked, a risk analysis was conducted prior to the study in order to comply with French (European) regulation. The clinical protocol was then authorized by the national agency for the safety of medicine and health products (ANSM, previously AFSSAPS) and approved by the relevant ethics committee (CPP Sud-Est V - Grenoble). This translational research is, to the best of our knowledge, the first study designed with high-level methodology to evaluate this innovative medical device in a comparative way using persons with SCI. This study may therefore be the first contribution towards estimating the efficiency of the proposed strategy to reduce prolonged excessive pressure in the buttock area.

TABLE IVA
ACCEPTABILITY OF THE PRESSURE MAP SENSOR FOR USE IN DAILY LIFE

| | Subject | SetUp | FacilityUse | ComfUse | PsyImp |
|---|---|---|---|---|---|
| Control group | 2 | 2 | 0 | 1 | 1 |
| | 6 | 0 | 0 | 0 | 0 |
| | 8 | 1 | 0 | 5 | 7 |
| | 9 | 1 | 1 | 2 | 8 |
| | 11 | 1 | 1 | 1 | 3 |
| | median | 1 | 0 | 1 | 3 |
| | [Q1; Q3] | [1; 1] | [0; 1] | [1; 2] | [1; 7] |
| Intervention group | 3 | 0 | 0 | 0 | 3 |
| | 4 | 1 | 0 | 3 | 0 |
| | 5 | 0 | 0 | 6 | 6 |
| | 7 | 0 | 0 | 0 | 0 |
| | median | 0 | 0 | 1.5 | 1.5 |
| | [Q1; Q3] | [0; 0.25] | [0; 0] | [0; 3.75] | [0; 3.75] |

Each aspect of the pressure map sensor was evaluated from 0 (acceptable) to 10 (unacceptable) for use in daily life.

TABLE IVB
ACCEPTABILITY OF THE TDU FOR USE IN DAILY LIFE

| Subject | SetUp | Remove | LossSens | ComfUse | PsyImp |
|---|---|---|---|---|---|
| 3 | 0 | 0 | 0 | 10 | 10 |
| 4 | 4 | 4 | 2 | 7 | 2 |
| 5 | 2 | 2 | 0 | 10 | 10 |
| 7 | 0 | 0 | 0 | 10 | 10 |
| median | 1 | 1 | 0 | 10 | 10 |
| [Q1; Q3] | [0; 2.5] | [0; 2.5] | [0; 0.5] | [9.25; 10] | [8; 10] |

Each aspect of the TDU was evaluated from 0 (acceptable) to 10 (unacceptable) for use in daily life.

Over two years, only twelve persons with SCI were recruited for this study, despite advertisements being published in the local press. The limited number of subjects recruited illustrates the difficulty of conducting such a study with persons with SCI involving a non CE-marked innovative medical device.

An interim analysis was planned after ten effectively analyzable subjects had been included with the stopping criterion being that the main outcome was statistically significant (p<0.01). The inclusion of an interim analysis with a stopping criterion is a standard aspect of clinical trial methodology [22,23]. Indeed, it is encouraged to stop the trial if evidence of benefit has been demonstrated [24]. For this study, the interim analysis was particularly important considering the difficulty of recruiting persons with SCI and the considerable participation required from the included subjects. However, an irreversible failure of the TDU occurred after twelve subjects had been included, of which only nine were effectively analyzable. The data of subject 1 (control group) was not exploitable because of a software failure (calibration reference maps not recorded), subject 10 was excluded due to an unexpected pregnancy (exclusion criteria, subject excluded before randomization) and subject 12 (intervention group) could not take part in the second session because of the irreversible failure of the TDU. The steering committee decided to conduct the statistical analysis at this point to evaluate whether the study could be prematurely stopped before considering whether to replace the TDU (approximate cost 50,000€). The ITT and PP statistical analysis was performed on respectively eleven and nine subjects. As the main outcome was statistically significant for both analyses (ITT: p=0.004; PP: p=0.008), the stopping criterion had therefore been met and the steering committee decided to stop the study.

The main outcome of this study shows that subjects in the intervention group were able to adopt adequate weight shifting when receiving electrostimulations of the tongue by the TDU (session 2). Although the TDU had been previously tested by one subject with paraplegia (in only two directions) [16], the results of this study demonstrate statistically, for the first time, that persons with SCI (i) successfully perceived the TDU electrostimulations, and (ii) were able to modify their posture according to the electrostimulation of their tongue. In the control group, the improvement in adequate weight shifting between the two sessions was null as expected. On the contrary, in the intervention group, as the information was delivered through the TDU during the second session, the median improvement in adequate weight shifting was 86%. However, the median proportion of adequate weight shifts in the second session (p2) of the intervention group was 0.89, rather than the theoretical score of 1 that had been expected. This could be due to a number of different reasons: (i) possible algorithm failure arising from the identification of the postures from the pressure maps [25]; (ii) hardware failure (TDU); (iii) lack or difficulty of perception of the electrostimulation of the tongue; (iv) incorrect interpretation of the TDU message [11] and (v) impossibility for the subject to follow the given instruction, that is, to move their trunk in the indicated direction.

One secondary outcome measured the evolution of an estimation of the risk of developing PU. In this study, no statistical difference was observed between the initial distribution of the risk index (R1) in the control and intervention group, indicating a similar risk index in both groups during session 1. Though no evolution was expected, the risk index doubled between session 1 and session 2 in the control group. We have no immediate explanation for such a behavior, though one possible reason is less stress and thus less movement during the second session. Considering the statistically significant PP analysis results, the complete opposite result was obtained in the intervention group: the risk index was on average halved when the TDU was activated. Furthermore, a ratio of approximately 4 between the control and intervention group illustrates the efficiency of such a strategy.

The ITT analysis of the evolution of the risk index in the intervention group was not statistically significant. This is due to subject 3 who displayed a different behavior to the overall group behavior in terms of the main outcome Δp (subject 3 = 0.24 << median intervention group = 0.86) and the first secondary outcome R2/R1 (subject 3 = 1.92 >> median intervention group = 0.55). This was explained by the pressure measured at one particular sensor $u_0$, located at coordinates {23,16} (approximate center of the left buttock), which contributed 49% of the risk index R2 measured during the second session. The highly localized nature of the excessive pressure (1 of 1024 sensors) makes it probable that it was caused by a very locally specific problem (temporary sensor dysfunction, a fold in the subjects clothing, a forgotten object in the rear pocket) rather than for example overall fatigue of the subject, which one would expect to impact a greater number of sensors.

Finally, as such a device would be used daily, a measure of its acceptability and usability was conducted. The results of the questionnaire indicated that the pressure map sensor of the device was positively accepted. However, the TDU of the device was very poorly rated. A different alerting device, acceptable for daily use by persons with SCI, therefore needs to be designed. Different options are currently being investigated such as, for example, the replacement of the TDU by a watch or a smartphone. Though these options are financially advantageous (much cheaper than the TDU), they could not be used by persons with tetraplegia, as the TDU can. A new design for the pressure map sensor part of the device is also currently being studied, in order to render it mobile with energetic autonomy.

Other improvements currently being studied involve work to optimize the alert threshold and the alert frequency. Both values were empirically defined for this study, as due to insufficient evidence, no definitive numerical values exist [26]. The alert threshold ($P_{risk}$) was defined as 100 mmHg, in the physiological range of arterial pressure. Pressure – the force per unit area - is a function of both the mass and the morphology of the subject. Though pressure increases with increasing mass, studies have measured higher pressure under the bony prominences of subjects classified as thin than subjects classified as obese due to the reduced surface area [27, 28]. The alert frequency was defined as 60 s for this study as it has been estimated that able bodied office workers change seated position approximately once per minute [17]. An important perspective is to optimize both alerts such that the system is acceptable for daily life while minimizing the risk of developing PU (alert only in the case of real danger).

The optimization of the alert threshold and frequency would enable the definition of the risk index to be improved. For this study, it was necessary to have a quantifiable means of estimating the risk of developing PU. As no such risk index, to the best of our knowledge, exists in literature, an index was defined that combines the two main factors that contribute to the risk of developing PU: the level and the duration of applied pressure, that are recognized as being correlated with PU [29]. Despite the evident limits of the defined risk index, consideration of the relative, rather than absolute, values of the index enabled the feasibility of using perceptive supplementation to aid subjects with SCI to reduce the risk of developing PU to be demonstrated.

A limit of this study is the small sample size which was caused by the irreparable failure of the device and the unjustifiably high cost to replace it considering that the device was evaluated as being very inconvenient for use in daily life. Nevertheless, this study has demonstrated with statistical significance (main outcome: p<0.01, PP and ITT) the feasibility of using perceptive supplementation to reduce prolonged excessive pressure in the buttock area and is a promising method to prevent PU. If the acceptability/usability issue can be resolved, such a device would be useful for (i) preventing PU in fragile persons with SCI, (ii) educating persons with recent SCI to adopt a safe behavior to prevent future PU, (3) preventing PU in other at-risk groups (for instance elderly, diabetic).

## V. Conclusion

We have reported here a study that evaluates the feasibility of an innovative medical device that aims to reduce prolonged excessive pressure, recognized as one of the main causes of pressure ulcers in persons with paraplegia. The device is composed of a pressure map sensor that measures the pressure at the buttock area in real time, combined with lingual tactile feedback that alerts and guides the user on how to adapt their posture in order to reduce prolonged excessive pressure.

The device was tested by eleven persons with paraplegia, separated in two groups: the control group and the intervention group. Global results show that the device enabled the subjects in the intervention group to reduce prolonged excessive pressure. Closer analysis of the results show that one subject in the intervention group did not benefit from the device, mainly due to the results of one particular sensor. Interestingly, our system could be adapted to detect this type of abnormality. In terms of acceptability, the subjects evaluated the TDU part of the device as being very inconvenient for use in daily life.

It is a perspective of this study to redesign the TDU part of the device such that it is acceptable for use in daily life (for example, using a watch or a smartphone). Further studies are also required to confirm these preliminary results.

## Acknowledgment

The authors gratefully acknowledge the contribution of each of the subjects included in the study. This study was made possible thanks to the TIMC-IMAG laboratory (Grenoble Alpes University, France) that made available the prototype of the evaluated medical device. The authors also acknowledge Guglielmi Technologies Dentaires who produced the wireless Tongue Display Unit for each of the study subjects and Vista Medical company who provided the pressure map sensor.

## References

[1] W. O. McKinley, A. B. Jackson, D. D. Cardenas, and J. Michael, "Long-term medical complications after traumatic spinal cord injury: a regional model systems analysis," *Arch. Phys. Med. Rehabil.*, vol. 80, no. 11, pp. 1402–1410, 1999.
[2] G. Bennett, C. Dealey, and J. Posnett, "The cost of pressure ulcers in the UK," *Age Ageing*, vol. 33, no. 3, pp. 230–235, 2004.
[3] E. Linder-Ganz, S. Engelberg, M. Scheinowitz, and A. Gefen, "Pressure–time cell death threshold for albino rat skeletal muscles as related to pressure sore biomechanics," *J. Biomech.*, vol. 39, no. 14, pp. 2725–2732, 2006.
[4] Consortium for Spinal Cord Medicine Clinical Practice Guidelines, "Pressure ulcer prevention and treatment following spinal cord injury: a clinical practice guideline for health-care professionals.," *J. Spinal Cord Med.*, vol. 24, p. S40, 2001.
[5] L. Stockton and S. Rithalia, "Is dynamic seating a modality worth considering in the prevention of pressure ulcers?," *J. Tissue Viability*, vol. 17, no. 1, pp. 15–21, 2008.
[6] M. Makhsous *et al.*, "Measuring tissue perfusion during pressure relief maneuvers: insights into preventing pressure ulcers," *J. Spinal Cord Med.*, vol. 30, no. 5, pp. 497–507, 2007.
[7] J. Reenalda, P. Van Geffen, G. Snoek, M. Jannink, M. Ijzerman, and H. Rietman, "Effects of dynamic sitting interventions on tissue oxygenation in individuals with spinal cord disorders," *Spinal Cord*, vol. 48, no. 4, p. 336, 2010.
[8] P. Bach-y-Rita, "Sensory plasticity. Applications to a vision substitution system," *Acta Neurol. Scand.*, vol. 43, no. 4, pp. 417–426, 1967.
[9] M. Tyler, Y. Danilov, and P. Bach-y-Rita, "Closing an open-loop control system: vestibular substitution through the tongue," *J. Integr. Neurosci.*, vol. 2, no. 02, pp. 159–164, 2003.
[10] P. Bach-y-Rita, K. A. Kaczmarek, M. E. Tyler, and J. Garcia-Lara, "Form perception with a 49-point electrotactile stimulus array on the tongue: a technical note," *J. Rehabil. Res. Dev.*, vol. 35, no. 4, p. 427, 1998.
[11] F. Robineau, F. Boy, J.-P. Orliaguet, J. Demongeot, and Y. Payan, "Guiding the surgical gesture using an electro-tactile stimulus array on the tongue: A feasibility study," *IEEE Trans. Biomed. Eng.*, vol. 54, no. 4, pp. 711–717, 2007.
[12] N. Vuillerme, O. Chenu, J. Demongeot, and Y. Payan, "Improving human ankle joint position sense using an artificial tongue-placed tactile biofeedback," *Neurosci. Lett.*, vol. 405, no. 1–2, pp. 19–23, 2006.
[13] N. Vuillerme, M. Boisgontier, O. Chenu, J. Demongeot, and Y. Payan, "Tongue-placed tactile biofeedback suppresses the deleterious effects of muscle fatigue on joint position sense at the ankle," *Exp. Brain Res.*, vol. 183, no. 2, pp. 235–240, 2007.
[14] K. A. Kaczmarek, "The tongue display unit (TDU) for electrotactile spatiotemporal pattern presentation," *Sci. Iran.*, vol. 18, no. 6, pp. 1476–1485, 2011.
[15] O. Chenu, N. Vuillerme, J. Demongeot, and Y. Payan, "A wireless lingual feedback device to reduce overpressures in seated posture: A feasibility study," *PLOS One*, vol. 4, no. 10, p. e7550, 2009.
[16] O. Chenu *et al.*, "Pressure sores prevention for paraplegic people: Effects of visual, auditory and tactile supplementations on overpressures distribution in seated posture," *Appl. Bionics Biomech.*, vol. 9, no. 1, pp. 61–67, 2012.
[17] A. Hedge and M. Ruder, "Dynamic sitting—how much do we move when working at a computer?," in *Proceedings of the Human Factors and Ergonomics Society Annual Meeting*, 2003, vol. 47, pp. 947–951.
[18] A. J. Elliot, *Handbook of approach and avoidance motivation*. Psychology Press, 2013.
[19] R Development Core Team, *R: A language and environment for statistical computing*. R foundation for statistical computing Vienna, Austria, 2008.
[20] W. S. El Masry, M. Tsubo, S. Katoh, Y. H. El Miligui, and A. Khan, "Validation of the American spinal injury association (ASIA) motor score and the national acute spinal cord injury study (NASCIS) motor score," *Spine*, vol. 21, no. 5, pp. 614–619, 1996.
[21] http://www.asia-spinalinjury.org/
[22] D. Moher *et al.*, "CONSORT 2010 explanation and elaboration: updated guidelines for reporting parallel group randomised trials," *Bmj*, vol. 340, p. c869, 2010.
[23] A. Kumar and B. S. Chakraborty, "Interim analysis: A rational approach of decision making in clinical trial," *J. Adv. Pharm. Technol. Res.*, vol. 7, no. 4, p. 118, 2016.
[24] J. Whitehead, "Stopping clinical trials by design," *Nat. Rev. Drug Discov.*, vol. 3, no. 11, p. 973, 2004.
[25] O. Chenu, "Conception et validation d'un dispositif de suppléance perceptive dédié à la prévention des escarres," Ph.D. dissertation, Université Joseph-Fourier-Grenoble I, 2009.
[26] S. Coleman *et al.*, "A new pressure ulcer conceptual framework," *J. Adv. Nurs.*, vol. 70, no. 10, pp. 2222–2234, 2014.
[27] T. W. Kernozek, P. A. Wilder, A. Amundson, and J. Hummer, "The effects of body mass index on peak seat-interface pressure of institutionalized elderly," *Arch. Phys. Med. Rehabil.*, vol. 83, no. 6, pp. 868–871, 2002.
[28] S. L. Garber and T. A. Krouskop, "Body build and its relationship to pressure distribution in the seated wheelchair patient.," *Arch. Phys. Med. Rehabil.*, vol. 63, no. 1, pp. 17–20, 1982.
[29] A. Gefen, B. van Nierop, D. L. Bader, and C. W. Oomens, "Strain-time cell-death threshold for skeletal muscle in a tissue-engineered model system for deep tissue injury," *J. Biomech.*, vol. 41, no. 9, pp. 2003–2012, 2008.